\documentstyle[12pt]{article}

\newcommand{\bear}{\begin{eqnarray}}
\newcommand{\eq}[1]{eq.~(\ref{#1})}
\newcommand{\eqs}[2]{eqs.(\ref{#1},\ref{#2})}

\newcommand{\eqsss}[2]{eqs.(\ref{#1}--\ref{#2})}
\newcommand{\Eq}[1]{Eq.~(\ref{#1})}

\newcommand{\beq}{\begin{equation}}
\newcommand{\eeq}{\end{equation}}

\newcommand{\la}[1]{\label{#1}}
\newcommand{\ear}{\end{eqnarray}}

\newcommand{\at}{\overline{10}}

  \newcommand{\bq}[1]{\begin{equation}{#1}\end{equation}}
  
  \def\beqr{\begin{eqnarray}}
  \def\eeqr{\end{eqnarray}}

  \def\dim{\mbox{dim}}
  \def\Tr{\mbox{Tr}}

  \def\bra#1{{\langle#1\vert}}
  \def\ket#1{{\vert#1\rangle}}
  
\def\appendix{\par
\setcounter{subsection}{0}
\setcounter{equation}{0}
\def\thesection{Appendix}
\def\theequation{\Alph{section}.\arabic{equation}}
}
\begin{document}
\vspace{2.5cm}

\thispagestyle{empty}
\rightline{RUB-TPII-02/97}
\rightline{NORDITA-97/19 N}
\rightline{hep-ph/9703373 }
\begin{center} {\Large\bf
Exotic Anti-Decuplet of Baryons: \\
\vspace{10pt}
Prediction from Chiral Solitons}\\
\vspace{2cm}
{\large\bf
Dmitri Diakonov$^{\diamond *}$, Victor Petrov$^\diamond$ and Maxim
Polyakov$^{\diamond\dagger}$}
\footnote{ \noindent e-mail:
maximp@hadron.tp2.ruhr-uni-bochum.de},\\

\vspace{40pt}
\noindent
{\small\it $^\diamond$Petersburg Nuclear Physics Institute, Gatchina,
St.Petersburg 188 350, Russia \\
$^*$NORDITA, Blegdamsvej 17, 2100 Copenhagen, Denmark \\
$^\dagger$Inst. f\"ur Theor. Physik I\hskip -1.5pt I, Ruhr-Universit\"at
Bochum, D-44780 Bochum, Germany}
\end{center}
\vspace{3.5cm}

\abstract{We predict an exotic $Z^+$ baryon
(having spin $1/2$, isospin $0$ and strangeness $+1$) with a relatively
low mass of about $1530\;MeV$ and total width of less than
$15\;MeV$.  It seems that this region of masses has avoided
thorough searches in the past.}
\newpage

\section{All light baryons are rotational excitations}

 The most striking success of the old Skyrme idea \cite{Skyrme} that
 nucleons can be viewed as solitons of the pion (or chiral) field,
 is the classification of light baryons it suggests. Indeed,
 the minimal generalization of spherical symmetry to incorporate
 three isospin components of the pion field is the so-called
 hedgehog form,

 \beq
 \pi^a(\vec{x})=\frac{x^a}{r}P(r),
 \eeq
 where $P(r)$ is the spherically-symmetric profile of the soliton. It
 implies that a space rotation of the field is equivalent to that in
 isospace. Hence, the quantization of the soliton rotation is similar
 to that of a spherical top: the rotational states have isospin
 $T$ equal to spin $J$, and their excitation energies are

 \beq
 E_{\mbox{rot}}=\frac{J(J+1)}{2I},
 \eeq
 where $I$ is the soliton moment of inertia. The rotational states are,
 therefore,
 $(2J+1)^2$-fold degenerate (in spin and isospin). For $J=1/2$ we have the
 four nucleon states; for $J=3/2$ we have the sixteen $\Delta$-isobar
 states.
 By saying that $N$ and $\Delta$ are different rotational states
 of the same object (the ``classical nucleon") one gets certain
 relations between their characteristics which are all satisfied up to
 a few percent in nature \cite{ANW}.

 Quantum Chromodynamics has shed some light into why the chiral soliton
 picture is correct: we know now that the spontaneous chiral symmetry
 breaking in QCD is, probably, the most important feature of strong
 interactions, determining to a great extent their dynamics
 (see, e.g. \cite{D}), while the large $N_c$
 (= numbers of colours) argumentation by Witten \cite{W}
 explains why the pion field inside the nucleon can be considered as a
 classical one, {\em i.e.} as a ``soliton".

 The generalization to hyperons \cite{W, Guad} makes the success of the
 chiral soliton idea even more impressive. The rotation can be now
 performed in the ordinary and in the flavour $SU(3)$ space.
 Its quantization shows
 \cite{Guad, DP, Prasz, Chemtob, ind}
 that the
 lowest baryon state is the {\em octet with spin $1/2$} and the next is the
 {\em decuplet with spin $3/2$} -- exactly what we meet in reality.
 Again, there are numerous relations between characteristics of members
 of octet and decuplet which follow purely from symmetry
 considerations. Perhaps the most spectacular is the Guadagnini formula
 \cite{Guad} which relates splittings inside the decuplet with those in
 the octet: it is satisfied to the accuracy better than one percent,
 see below.

 What are the next rotational excitations? If one restricts oneself to
 only two
 flavours, the next state should be a (5/2, 5/2) resonance; in the
 three-flavour case the third rotational excitation is an
 {\em anti-decuplet with spin $1/2$} \footnote{Probably the existence of the
 anti-decuplet as the next $SU(3)$ rotational excitation has been
 first pointed out by
 the authors at the ITEP Winter School (February, 1984), see ref.
 \cite{DP}.  Other early references for the anti-decuplet include
 \cite{Chemtob, Prasz2, Walliser}.}.  Why do not we have any clear
 signal of the exotic (5/2, 5/2) resonance?  The reason is that the
 angular momentum $J=5/2$ is numerically comparable to $N_c=3$.
 Rotations with $J\sim N_c$ cannot be considered as slow:  the
 centrifugal forces deform considerably the spherically-symmetric
 profile of the soliton field \cite{DPRegge, BR}; simultaneously at $J\sim N_c$ the
 radiations of pions by the rotating body makes the total width of the
 state comparable to its mass \cite{DPRegge,Dorey}. In order to
 survive strong pion radiation the rotating chiral solitons with $J\ge
 N_c$ have to stretch into cigar-like objects; such states lie on
 linear Regge trajectories \cite{DPRegge}.

 The situation, however, might be somewhat different in the
 {\it three}-flavour case. First, the rotation is, roughly
 speaking, distributed among more axes in flavour space, hence
 individual angular velocities are not neccessarily as large as when we
 consider the two-flavour case with $J=5/2$. Actually, the $SU(2)$
 baryons with $J=5/2$ belong to a very high multiplet from the $SU(3)$
 point of view. Second, the radiation by the soliton includes now $K$
 and $\eta$ mesons which are substantially heavier than pions, and
 hence such radiation is to some extent suppressed. Therefore, the
 anti-decuplet baryons may not neccessarily have widths comparable to
 masses. And this is what we, indeed, find below.

 We conclude thus that an expectaction of a relatively light and narrow
 anti-decuplet of baryons is theoretically motivated. Moreover, we are
 in a position to numerically estimate the masses and widths of the
 members of the anti-decuplet, and to point out possible experiments to
 observe them.

 Let us mention that one can altogether abandon the soliton logic:
 the exotic anti-decuplet can be alternatively considered in a primitive
 way as states made of three
 quarks plus a quark-antiquark pair, or else, as a bound state of octet
 baryons with octet mesons. For example, the most interesting member of
 the anti-decuplet, viz. the exotic $Z^+$ baryon can be considered as a
 $K^+ n$ or $K^0 p$ bound state \footnote{It is known that the
 $KN$ phase shift in the $T=0$, $J=(1/2)^+$ state
 corresponds to attraction \cite{Dover}.}.
 Unfortunately, the predictions become
 then to a great extent model-dependent. It is a big advantage of the
 chiral soliton picture that all concrete numbers (for masses and
 widths) do not rely upon a specific dynamical realization but follow
 from symmetry considerations. Actually, only one number would be
 useful to get from dynamics, namely a specific $SU(3)$ moment of
 inertia $I_2$ (see below), and concrete dynamical models give concrete values
 for this quantity. In this paper, however, we prefer to extract this
 quantity from experiment -- by identifying the known nucleon resonance
 $N\left(1710, \frac 12^+\right)$ with the member of the suggested
 anti-decuplet \footnote{A possibility for such an identification has
 been mentioned in ref. \cite{SW} but other members of the
 anti-decuplet have not been addressed there.}.
 We, then, are able to fix completely all the other members of the
 anti-decuplet together with their widths and branching ratios.

 To end up this introduction we draw the $SU(3)$ diagramm for the
 suggested anti-decuplet in the $(T_3,Y)$ axes, indicating its naive
 quark content as well as the (octet baryon + octet
 meson) content, see Fig.~1. In addition to the lightest
 $Z^+$ there is an exotic quadruplet of $S=-2$ baryons (we call them
 $\Xi_{3/2}$). However, the exotic $\Xi^{--}$ and $\Xi^+$ hyperons
 appear to be very heavy and to have large widths, and can
 therefore hardly be detected.  Therefore, apart from peculiar
 branching ratios predicted for the $N\left(1710, \frac 12^+\right)$
 and the $\Sigma\left(1880, \frac{1}{2}^+\right)$ resonances, the
 crucial prediction is the existence of a relatively light and narrow
 $Z^+$ baryon.

\section{Rotational states}

Following Witten \cite{W} and Guadagnini \cite{Guad} we assume the
self-consistent pseudoscalar field which binds up the $N_c = 3$ quarks
in the ``classical'' baryon (i.e. the soliton field) to be of the
form

\beq
U(\vec{x})\equiv \exp\left(i \pi_A(\vec{x})\lambda^A/F_\pi\right)
= \left(\begin{array}{cc}
\exp \left[i(\mbox{\boldmath{$n$}}
\cdot\mbox{\boldmath{$\tau$}}) P(r)\right] &
\begin{array}{c} 0\\ 0\end{array}\\ \begin{array}{cc}0 &0 \end{array} & 1
\end{array}\right),\;\;\;\; \vec{n}=\frac{\vec{x}}{r},
\la{hedgehog}
\eeq
where the spherically-symmetric profile function $P(r)$ is defined by
dynamics. We shall not need the concrete form of this function in what
follows. In \eq{hedgehog} $\lambda^A$ are the eight Gell-Mann $SU(3)$
matrices, and $\mbox{\boldmath{$\tau$}}$ are the three Pauli $SU(2)$
matrices.

In order to provide the ``classical'' baryon with specific quantum
numbers one has to consider a $SU(3)$-rotated pseudoscalar field,

\beq
\tilde{U}(\vec{x},t)=R(t)U(\vec{x})R^+(t)
\la{rothedgehog}
\eeq
where $R(t)$ is a unitary $SU(3)$ matrix depending only on time and
$U(\vec{x})$
is the static hedgehog field given by \eq{hedgehog}.
Quantizing this rotation one gets the following rotational
Hamiltonian \cite{Guad},

\beq
H^{rot}=\frac{1}{2I_1}\sum_{A=1}^3
J_A^2+\frac{1}{2I_2} \sum_{A=4}^7 J_A^2,
\la{Ham1}
\eeq
where $J_A$ are the generators of the $SU(3)$ group; $J_A$ with $A=1,2,3$
are the usual angular momentum (spin) operators, and $I_{1,2}$ are the
two $SU(3)$ moments of inertia, which are model-dependent.
Most important is the additional quantization prescription,

\beq
J_8=-\frac{N_cB}{2\surd{3}}=-\frac{\surd{3}}{2},\;\;\;\;\;
Y^\prime=-\frac{2}{\surd{3}}J_8=1,
\la{quantiz}
\eeq
where $B$ is the baryon number, $B=1$.
In the Skyrme model this quantization rule follows from the Wess-Zumino
term \cite{W,Guad}. In the more realistic chiral quark--soliton model
\cite{DPP} it arises from filling in the discrete level, i.e. from
the ``valence" quarks \cite{mass}. It is known to lead to the selection
rule
\cite{Guad,DP,Prasz,Chemtob,ind}:
not all possible spin and $SU(3)$
multiplets are allowed as rotational excitations of the $SU(2)$
hedgehog.  \Eq{quantiz} means that only those $SU(3)$ multiplets are
allowed which contain particles with hypercharge $Y=1$; if the number
of particles with $Y=1$ is denoted as $2J+1$, the spin of the allowed $SU(3)$
multiplet is equal to $J$.

Therefore, the lowest allowed $SU(3)$ multiplets are:

\begin{itemize}
\item octet with spin 1/2 (since there are {\em two} baryons in the octet
with $Y=1$, the $N$)
\item decuplet with spin 3/2 (since there are {\em four} baryons in the
decuplet with $Y=1$, the $\Delta$)
\item anti-decuplet with spin 1/2 (since there are {\em two} baryons in the
anti-decuplet with $Y=1$, the $N^*$)
\end{itemize}
The next are 27-plets with spin $1/2$ and $3/2$ but we do not
consider them here. The appropriate rotational wave functions describing
members of these multiplets are given in Appendix A.

For the representation $(p,q)$ of the $SU(3)$ group one has

\beq
\sum\limits_{A=1}^8 J_A^2 = \frac13[p^2+q^2+pq + 3(p+q)],
\eeq
therefore the eigenvalues of the rotational Hamiltonian
(\ref{Ham1}) are

\beq
E_{(p,q)}^{rot} = \frac1{6I_2} [p^2+q^2+pq + 3(p+q)]
+\left(\frac1{2I_1} - \frac1{2I_2} \right) J(J+1)
- \frac{(N_c B)^2}{24I_2}.
\la{rotenerg}\eeq
We have the following three lowest rotational excitations:

\beqr
(p,q)=(1,1), & J=1/2& \mbox{(octet)} \\
(p,q)=(3,0), & J=3/2& \mbox{(decuplet)} \\
(p,q)=(0,3), & J=1/2& \mbox{(anti-decuplet)} \, .
\eeqr
The splittings between the centers of these multiplets are determined by the
moments of inertia, $I_{1,2}$:

\beq
\Delta_{10-8} =
E_{(3,0)}^{rot} - E_{(1,1)}^{rot} =\frac{3}{2I_1},
\la{octdecspl}\eeq
\beq
\Delta_{{\overline{10}} - 8} =
E_{(0,3)}^{rot} - E_{(1,1)}^{rot} =\frac{3}{2I_2},
\la{oct-antidecspl}\eeq
\beq
\Delta_{{\overline{10}} - 10} =
E_{(0,3)}^{rot} - E_{(3,0)}^{rot} =
\frac{3}{2I_2} - \frac{3}{2I_1}.
\la{dec-antidecspl}\eeq

We see that, were the moment of inertia $I_2 > I_1$, the anti-decuplet
would be even lighter than the standard decuplet. Though we do not
know of any strict theorem prohibiting this inequality, all
 dynamical models we know of give $I_1 > I_2$, so that the
 anti-decuplet appears to be heavier.

\section{Splittings in the $SU(3)$ multiplets}

We now take into account the non-vanishing strange quark mass.
The effects of the non-zero $m_s$ are of two kind: first, it splits
the otherwise degenerate masses inside each $SU(3)$ multiplet; second,
it leads to mixing between different $SU(3)$ multiplets.
We shall systematically restrict ourselves to the linear order in $m_s$.
In this order the phenomenologically successful Gell-Mann--Okubo
and Guadagnini formulae are automatically satisfied.

Theoretically, the corrections to baryon masses due to $m_s\neq 0$ are
of two types: i)~leading order, $O(m_sN_c)$ and ii) subleading order,
$O(m_sN_c^0)$.  These corrections perturb the rotational Hamiltonian
(\ref{Ham1}) (for derivation see ref.~\cite{mass}) by

\beq
\Delta H_m= \alpha D_{88}^{(8)} + \beta Y
+ \frac{\gamma}{\surd 3} \sum\limits_{i=1}^3 D_{8i}^{(8)} J_i \, ,
\label{Ham2} \eeq
where $D_{...}^{(8)}(R)$ are the Wigner $SU(3)$ finite-rotation
matrices depending on the orientation matrix of a baryon, see the Appendices.
The coefficients $\alpha, \beta, \gamma$ are proportional to the
mass of the $s$ quark and are expressed through a combination of
the soliton moments of inertia, $I_{1,2}$ and $K_{1,2}$, and the
nucleon sigma term, $\Sigma$ \cite{mass}:

\beq
\alpha = - \, \frac23 \, \frac{m_s}{m_u + m_d} \, \Sigma
+ m_s \frac{K_2}{I_2} \, ,
\la{a}\eeq
\beq
\beta = - m_s \frac{K_2}{I_2} \, ,
\la{b}\eeq
\beq
\gamma = \frac{2}{3} m_s\left( \frac{K_1}{I_1}
- \frac{K_2}{I_2} \right) \,,
\la{g}\eeq
\beq
\Sigma = \frac{m_u+m_d}{2}\langle N|\bar u u+\bar d d|N\rangle\,.
\la{s}\eeq

To get the physical splittings from \eq{Ham2} one has to sandwich it
between the physical rotational states:

\bq{ \Delta m_B = \bra{B} \Delta H_m \ket{B}. }
The mass splittings inside multiplets in terms of the coefficients
$\alpha, \beta, \gamma$ are
listed in Table~1~\footnote{We take the opportunity to thank
M.Praszalowicz and P.Pobylitsa who have
participated in calculating this table back in 1988.}. We call the two
members of the
anti-decuplet with exotic quantum numbers $T=0,\, S=1$ and $T=3/2,\,
S=-2$ as $Z^+$ and $\Xi_{3/2}$, respectively.

\vskip 0.33cm
\begin{tabular}{|c|c|c||c|}
\hline
\hline
octet & $T$ & $Y$ &
$\Delta m_B$ \\
\hline
$N$ & $1/2$ & $1$ &$(3 / 10)\alpha+\beta-(1/20)\gamma$\\
$\Lambda$ & $0$ & $0$& $(1/10)\alpha +(3/20)\gamma$\\
$\Sigma $ & $1$ & $0$& $-(1/10)\alpha -(3/20)\gamma$\\
$\Xi$ & $1/2$ & $-1$ &$-(1/5)\alpha-\beta+(1/5)\gamma $\\
\hline
\hline
decuplet & & & \\
\hline
$\Delta$ & $3/2$ & $1$ & $(1/8) \alpha + \beta - (5/16) \gamma$\\
$\Sigma^\ast$&$1$& $0$ & $0 $\\
$\Xi^\ast$ & $1/2$ & $-1$& $-(1/8)\alpha -\beta +(5/16) \gamma$\\
$\Omega$ & $0$ & $-2$& $-(1/4) \alpha -2\beta + (5/8) \gamma$\\
\hline
\hline
antidecuplet & & & \\
\hline
$Z^+$&$0$ &$2$ & $(1/4) \alpha +2\beta - (1/8) \gamma$\\
$N_{\at} $&$1/2$&$1$ & $(1/8) \alpha +\beta -(1/16) \gamma$\\
$\Sigma_{\at}$&$1$ &$0$ & $ 0 $\\
$\Xi_{3/2}$&$3/2$&$-1$ &$ -(1/8)\alpha - \beta + (1/16)\gamma$\\
\hline \hline
\end{tabular} \vskip 0.5cm

Table 1. Mass splittings within multiplets,
$\Delta m_B = \bra{B} \Delta H_m \ket{B}$.
\vskip 0.33cm

One observes that all splittings inside the octet and the decuplet
are expressed through only two combinations of $\alpha,\,\beta$ and
$\gamma$. This is the reason why, in the soliton picture, in addition
to the standard Gell-Mann--Okubo relations,

\beq
2(m_N+m_\Xi)=3m_\Lambda+m_\Sigma,
\la{GMOo}\eeq

\beq
m_\Delta-m_{\Sigma^*}=m_{\Sigma^*}-m_{\Xi^*}=m_{\Xi^*}-m_{\Omega^-},
\la{GMOd}\eeq
there arises a relation between the splittings inside the octet and
the decuplet, the Guadagnini formula \cite{Guad},

\beq
8(m_{\Xi^*}+m_N)+3m_\Sigma=11m_\Lambda+8m_{\Sigma^*},
\la{Gua}\eeq
which is satisfied with better than one-percent accuracy!
The best fit to the splittings in the octet and the decuplet gives
the following numerical values for the two combinations of the
coefficients $\alpha,\,\beta$ and $\gamma$:

\[
\alpha+\frac{3}{2}\gamma = -380\,MeV,
\]
\beq
\frac{1}{8}\alpha+\beta-\frac{5}{16}\gamma=-150 \,MeV.
\la{abg}\eeq

To learn the splittings in the anti-decuplet one needs to
know the third combination of $\alpha,\,\beta$ and $\gamma$, which is
not directly deducible from the octet and decuplet splittings. The only
statement which can be immediatelly made from looking into Table~1 is
that the spectrum in the anti-decuplet is equidistant, as in the normal
decuplet. The third combination can be fixed, however, from the
knowledge of the nucleon sigma term \cite{GLS} and of the current
quark mass ratio \cite{Leut}:

\beq
\frac{m_s}{m_u+m_d}\approx 12.5,\;\;\;\;\;\Sigma\approx 45\,MeV.
\la{massratio}\eeq
These numbers gives for the sum:

\beq
\alpha + \beta = -\frac{2 m_s}{3 (m_u+m_d)}\Sigma\approx -375\,MeV.
\la{ab}\eeq
Combining this knowledge with \eq{abg} we get finally all three
coefficients:

\beq
\alpha\approx -218\,MeV,\;\;\beta\approx -156\,MeV,\;\;
\gamma\approx -107\,MeV.
\la{abgnum}\eeq
The equidistant splittings inside the anti-decuplet are
thus expected to be equal to

\beq
\Delta m_{\at}=-\frac{\alpha}{8}-\beta+\frac{\gamma}{16}
\approx 180\,MeV\,,
\la{antispl}\eeq
the lightest baryon being the exotic $Z^+$ resonance.

To end up this section we note that the non-zero strange quark mass
leads also to the mixing of octet and anti-decuplet states with otherwise
identical quantum numbers.
In the linear order in $m_s$ these mixings are derived in Appendix A
and can be all expressed through one constant which we call $c_{\at}$,
where

\beq
c_{\at}=-\,\frac{1}{3\surd{5}}\left(\alpha+\frac{1}{2}\gamma
\right)I_2\,.
\la{smesh}\eeq

The true hyperon states become superpositions of the octet and
anti-decuplet states:

\vspace{1cm}
\underline{{\it Mainly octet baryons}}

\bear
|N\rangle&=& |N,8\rangle +c_{\at}|N,\at\rangle,\la{first}\\
|\Lambda\rangle&=& |\Lambda,8\rangle ,\\
|\Sigma\rangle&=& |\Sigma,8\rangle +c_{\at}|\Sigma,\at\rangle,\\
|\Xi\rangle&=& |\Xi,8\rangle ;
\ear

\underline{{\it Mainly anti-decuplet baryons}}

\bear
|Z^+\rangle&=& |Z^+,\at\rangle ,\\
|N_{\at}\rangle&=& |N,\at\rangle -c_{\at}|N,8\rangle,\\
|\Sigma_{\at}\rangle&=& |\Sigma,\at\rangle -c_{\at}|\Sigma,8\rangle,\\
|\Xi_{3/2}\rangle&=& |\Xi_{3/2},\at\rangle ,
\la{last}\ear

In the linear order in $m_s$ the mixing does not effect the mass
splittings inside the multiplets, discussed above.

Apart from $\alpha$ and $\gamma$, which we know now, the mixing
coefficient $c_{\at}$ is proportional to the second moment of inertia
$I_2$ which defines the shift of the anti-decuplet center (i.e. the
$\Sigma_{\at}$ baryon) from the octet center, see \eq{oct-antidecspl}.
We do not know of any symmetry considerations relating
this shift to that between the centers of the octet and the decuplet.
The dynamical (i.e. model) predictions for the moment
of inertia $I_2$ are rather disperse: they range from $0.43$~fm in the
Skyrme model \cite{Prasz2, Walliser}
to $0.55$~fm in the chiral quark--soliton model \cite{mass}. Taken
literally, the last value of $I_2$ would lead to a very light
anti-decuplet, and in particular to a $Z^+$ lying below the $KN$
threshold and thus stable against strong interactions. However,
it should be mentioned that the moments of inertia have $\sim
m_s$ corrections which are not computed yet. On physical grounds
one can argue that the $m_s$ corrections should be negative, since
the account for non-zero quark mass makes the baryons more ``tight".

In any case, we prefer to fix this
fundamental quantity from identifying one of the members of the
anti-decuplet, namely the one with the nucleon quantum numbers,
$N_{\at}$, with the rather well established nucleon resonance
$N\left(1710, \frac{1}{2}^+\right)$.
Given that $N_{\at}$ is $\approx 180\,MeV$ lighter than the center
of the anti-decuplet, we find

\beq
I_2 \approx 0.4\, \mbox{fm}\approx (500\,MeV)^{-1},
\la{I2}\eeq
and hence the octet--anti-decuplet mixing amplitude is

\beq
c_{\at}\approx 0.084\,,
\la{mixnum}\eeq
being not a negligible quantity.

We thus arrive to the following masses of the anti-decuplet:

\[
m_{Z^+} \approx 1530\, MeV,
\]
\[
m_{N_{\at}} \approx 1710\, MeV {\rm (input)},
\]
\[
m_{\Sigma_{\at}} \approx 1890\, MeV,
\]
\beq
m_{\Xi_{3/2}} \approx 2070\,MeV.
\la{masses}\eeq

\section{Baryon decays}

In the non-relativistic limit for the initial and final baryons the
baryon-baryon-meson coupling can be written in terms of rotational
coordinates $R$ of the baryon as \cite{ANW}

\beq
-i \frac{3 G_0}{2 m_B}\cdot \frac 12 \Tr(R^\dagger\lambda^m
R\lambda_i)\cdot p_i,
\la{meson-soliton}\eeq
where $\lambda^m$ is the Gell-Mann matrix for the emitted meson of flavour $m$,
and $p_i$ is the 3-momentum of the meson.
To make \eq{meson-soliton} more symmetric we use for $m_B$ in the denominator
the half-sum of the initial ($B_1$) and final ($B_2$) baryon masses.
Sandwiching
eq.~(\ref{meson-soliton}) between the rotational wave functions
of initial and final baryons, given by \eqs{wave-functions}{perth},
we obtain the $B_1\rightarrow B_2+M$ transitions
amplitude squared (averaged over the initial and summed over the final
spin and isospin states) in terms of the $SU(3)$ isoscalar factors. The
general formula is given in Appendix~B. Using it we get for
the particular modes of the $10\rightarrow 8 + 8$ decays:
\vspace{0.5cm}

\underline{{\it Decays of the decuplet}}

\bear
\Gamma(\Delta \rightarrow N\pi)&=&
 \frac{3 G_0^2}{2\pi (M_\Delta+M_N)^2} |\vec{p}|^3
\frac{M_N}{M_\Delta}\cdot\frac 15
= 110\,MeV\;vs.\;110\,MeV\;(exp.),
\label{ww1}
 \\
\Gamma(\Sigma^* \rightarrow \Lambda\pi)&=&
\frac{3 G_0^2}{2\pi (M_{\Sigma^*}+M_\Lambda)^2}
|\vec{p}|^3 \frac{M_\Lambda}{M_{\Sigma^*}}\cdot\frac{1}{10}
= 35\,MeV\;vs.\;35\,MeV\;(exp.),
\\
\Gamma(\Sigma^* \rightarrow \Sigma\pi)&=&
\frac{3 G_0^2}{2\pi (M_{\Sigma^*}+M_\Sigma)^2}
|\vec{p}|^3 \frac{M_\Sigma}{M_{\Sigma^*}}\cdot\frac{1}{15}
=5.3\,MeV\;vs.\;4.8\,MeV\;(exp.), \\
 \Gamma(\Xi^* \rightarrow \Xi\pi)&=&
\frac{3 G_0^2}{2\pi (M_{\Xi^*}+M_\Xi)^2} |\vec{p}|^3
\frac{M_\Xi}{M_{\Xi^*}}\cdot\frac{1}{10}
=8.6\,MeV\;vs.\;10\,MeV\;(exp.),
\label{ww4}
\ear
where $|\vec{p}|=\sqrt{(M_1^2-(M_2+m)^2)\cdot (M_1^2-(M_2-m)^2)}/2M_1$
is the momentum of the meson of mass $m$. To get the concrete numbers we
have used the value of the dimensionless coupling constant in
\eq{meson-soliton} $G_0\approx 19$.

We remind the reader that the usual $SU(3)$ symmetry would require {\em two}
coupling constants ($F$ and $D$) to determine the above widths, and of course
the $SU(3)$ symmetry by itself tells nothing about the relation between
decay constants for {\em different} multiplets. The chiral soliton models,
while preserving the usual $SU(3)$ symmetry, in addition give relations
between various couplings, since they all correspond to different rotation
states of the same object. In particular, the chiral soliton models
predict the $F/D$ ratio to be \cite{Pol}

\beq
\frac{F}{D}=\frac{5}{9}=0.555...\;\;\;\;vs.\;\;\;\;0.56\pm 0.02\;
({\rm exper.}),
\la{FD}\eeq
and the $g_{\pi NN}$ constant to be

\beq
g_{\pi NN}=\frac{7}{10}\,G_0\approx 13.3\;\;\;\;vs.\;\;\;\;\approx
13.6\; ({\rm exper.}).  \la{gpinn}\eeq

Again, we see that the notion of `baryons as rotational excitations'
works quite satisfactory. Therefore, one would expect that {\em the same}
coupling constant $G_0$ should be used for predicting the partial decay
rates of the next rotational excitation, the anti-decuplet.

We remark, however, that in the particular case of the pseudoscalar
couplings we expect rather large $1/N_c$ corrections which need not be
universal for all multiplets. The point is, the baryon-pseudoscalar
couplings are related, thanks to Goldberger--Treiman, to the baryon
axial constants, $g_A$. Meanwhile, it is well known that the real-world
($N_c=3$) value of the nucleon axial constant
$g_A$ differs from its large-$N_c$ limit roughly by a
factor \cite{BR2} $(N_c+2)/N_c=5/3$, which is quite significant.
This value comes from an estimate in a non-relativistic quark model and
is not necessarily exactly true, however it gives an idea of the size of the
$1/N_c$ corrections to the pseudoscalar couplings. Therefore, in order
to perform a reliable estimate of the anti-decuplet widths we have to
take into account, in addition to the leading-order \eq{meson-soliton},
the $1/N_c$ corrections to that formula. The relevant $1/N_c$ corrections
have been treated in ref. \cite{Christov} for the $SU(2)$ case and in
ref. \cite{BPG} for the $SU(3)$ octet and decuplet cases; below we extend
these works to the anti-decuplet couplings.

In the next-to-leading order one has to add to
eq.~(\ref{meson-soliton}) new operators depending on the
angular momentum $J_a$. These operators have the form
\cite{BPG}:

\beq
i \frac{3 G_1}{2 m_B}\cdot
d_{iab} \cdot \frac 12
\Tr(R^\dagger\lambda^m
R\lambda^a) J_b \cdot p_i +
i \frac{3 G_2}{2 m_B\sqrt 3}\cdot \frac 12
\Tr(R^\dagger\lambda^m
R\lambda^8) J_i \cdot p_i ,
\la{meson-soliton-rot}\eeq
where $d_{abc}$ is the $SU(3)$ symmetric tensor, $a,
b=4,5,6,7$, and $J_a$ are the generators of the infinitesimal $SU(3)$
rotations. The new coupling constants $G_{1,2}$ are suppressed by
$1/N_c$ relative to the leading-order coupling constant $G_0$,
although numerically they can be sizable.

Sandwiching eqs.~(\ref{meson-soliton},\ref{meson-soliton-rot}) between
the rotational wave functions of initial and final baryons and taking
into account the anti-decuplet--octet mixing represented by the
coefficient $c_{\at}$ (see see \eq{mixnum}), we obtain the following
general formula for partial widths of members of the decuplet and
of the anti-decuplet:

\bear
\Gamma(B_1 \rightarrow B_2 M)&=&
 \frac{3 G_r^2}{2\pi (M_1+M_2)^2} |\vec{p}|^3
\frac{M_2}{M_1}\cdot (C_1 + \frac{1}{\sqrt 5} C_2\cdot c_{\at}),
\label{wg}
\ear
The effective coupling constant $G_r$ depending on the multiplet and the
isoscalar factors $C_1$ and $C_2$ for various decays are listed in Table~2. The
pion nucleon
coupling constant $g_{\pi NN}$ and the $F/D$ ratio can be also
expressed in terms of the couplings $G_{0,1,2}$:

\bear
g_{\pi NN}&=& \frac{7}{10}\cdot (G_0 +\frac 12 G_1 +\frac{1}{14} G_2),\\
\frac{F}{D}&=& \frac 59\cdot \frac{G_0+\frac 12 G_1+\frac 12  G_2}
{G_0+\frac 12 G_1-\frac 16 G_2}\,.
\ear
Substituting in the last equation the experimental value of $F/D = 0.56\pm
0.02$ we find the value of the ratio:

\beq
\frac{G_2}{G_0+\frac 12 G_1}= 0.01\pm 0.05,
\label{g2num}
\eeq
which turns to be very small and will be neglected.
The smallness of $G_2$ is not surprising as it can be related to the singlet
axial constant of the nucleon,

\beq
G_2=\frac{2 M_N}{3 F_\pi} g_A^{(0)}\,,
\eeq
the latter known to be small. We see here another remarkable
prediction of the `baryon as soliton' idea: the smallness of
the singlet axial constant $g_A^{(0)}$ is directly related to
the smallness of the deviation of the $F/D$ ratio from 5/9.

\vskip 0.33cm
\begin{tabular}{|c|c|c|c|}
\hline
\hline
decay & $G_r$ & $C_1$ & $C_2$\\
\hline
\hline
$\Delta \to N\pi$ & $G_0+\frac 12 G_1$ &$\frac 15$ & 0 \\
$\Sigma^* \to \Lambda\pi$ & $G_0+\frac 12 G_1$ &$\frac{1}{10}$ & 0 \\
$\Sigma^* \to \Sigma\pi$ & $G_0+\frac 12 G_1$ &$\frac{1}{15}$ & 0 \\
$\Xi^*\to \Xi \pi $ & $G_0+\frac 12 G_1$ &$\frac{1}{10}$ & 0 \\
\hline
$N_{\at} \to N\pi$ & $G_0-G_1-\frac 12 G_2$ &
$\frac{1}{20}$ &$ -\frac{23}{40}$ \\
$N_{\at} \to N\eta$ & $G_0-G_1-\frac 12 G_2$ &
$\frac{1}{20}$ & $-\frac{1}{40} $\\
$N_{\at} \to \Delta\pi$ & $G_0 + \frac 1 2 G_1 $ &
0 & $\frac 45 c_{\at}$ \\
$N_{\at} \to \Lambda K$ & $G_0-G_1-\frac 12 G_2$ &
$\frac{1}{20}$ & $\frac{2}{ 5}$ \\
$N_{\at} \to \Sigma K$ & $G_0-G_1-\frac 12 G_2$ &
$\frac{1}{20}$ & $\frac{3}{20}$ \\
\hline
$ Z^+\to N K$ & $G_0-G_1-\frac 12 G_2$ &
$\frac{1}{5}$ & $\frac{1}{4}$ \\
\hline \hline
\end{tabular} \vskip 0.5cm

Table 2. Clebsch-Gordan coefficients entering eq.~(\ref{wg}) for the
decays of the decuplet and of the lightest members of the anti-decuplet.
\vskip 0.33cm

From Table~2 we see that the decuplet-octet couplings  are
proportional to $(G_0+G_1/2)$ and hence (if one neglects the
apparently small $G_2$) it can be related to the
pion-nucleon coupling $g_{\pi NN}$. This observation
means that the widths of the decuplet calculated in the leading $1/N_c$ order
in the beginning of this section are actually not affected by the
rotational $1/N_c$ corrections: in the next-to-leading order the
relation of the decuplet widths to the $g_{\pi NN}$ constant is not
changed. One has just to replace the $G_0$ of \eqsss{ww1}{gpinn} by
$G_0+G_1/2$. Therefore, from the decuplet decay width one finds
(cf. \eq{gpinn})

\beq
G_{10}=G_0+\frac 12 G_1 \approx 19,\;\;\;\;\;
g_{\pi NN} \approx \frac{7}{10}
\left(G_0+\frac{1}{2}G_1\right)\approx 13.3\,.
\eeq

However, the situation is different for the anti-decuplet: as seen
from Table~2, its effective couplings are proportional to $(G_0-G_1)$
(again we neglect the small $G_2$), not to $(G_0+\frac 12 G_1)$.
Therefore, with the $1/N_c$ corrections taken into account,
the anti-decuplet--octet coupling is not expressed solely through
$g_{\pi NN}$: to calculate the anti-decuplet decay widths one has
to know the ratio $G_1/G_0$ as well. Unfortunately, it can not be fixed in a
model-independent way -- one has to resort to some model. In
the chiral quark-soliton model \cite{DPP} the ratio $G_1/G_0$ is in
range from $0.4$ to $0.6$ \cite{Christov,BPG} \footnote{M.P. is
grateful to H.-C.~Kim and T.~Watabe for a detailed discussion of this
issue}. A similar calculation of the $G_2$ coupling in the same model
shows that it is substantially smaller than $G_0$ \cite{BPG},
in accordance with the experiment, see \eq{g2num}.
In the estimates below we use the lower value of the ratio,
$G_1/G_0\approx 0.4$, corresponding to the value of the anti-decuplet
decay constant,

\beq
G_{\at} \approx G_0-G_1 \approx 0.5\cdot G_{10} \approx 9.5\; .
\la{g1g0}\eeq

It should be mentioned that the non-relativistic quark model
(which, to some extent, can be used as a guiding line) predicts
$G_1/G_0=4/5$ and $G_2/G_0=2/5$, which is in a qualitative agreement
with a more realistic calculation in the quark soliton model.
Amusingly, though, these ratios produce exactly zero $G_{\at}$.
At the moment we are unable to point out the deep reason for
such a cancellation; in any case the non-relativistic quark
model cannot be considered as realistic as it gives also a too large
value of the $F/D$ ratio and of the singlet axial constant.
However, it may indicate that \eq{g1g0} over-estimates $G_{\at}$,
and that the widths of the anti-decuplet are even more narrow
than we estimate below.

We now present the decay rates of the members of the anti-decuplet
using \eqs{wg}{g1g0} with the Clebsch--Gordan coefficients from Table~2 which
also takes into account the octet--anti-decuplet mixing discussed in section~3.

\vspace{0.5cm}
\underline{$T=0,\; S=1$ state (the exotic $Z^+$ baryon) }

\beq
\Gamma(Z^+ \rightarrow N K)=
\frac{3 G_{\at}^2}{2\pi (M_N+M_{Z})^2} |\vec{p}|^3 \frac{M_N}{M_{Z}}
\cdot\frac{1}{5}\bigl(1+\frac{\sqrt 5}{4} c_{\at}\bigr)
=15\,MeV,
\label{zw}
\eeq
Since there are no other strong decay modes, the total width of the $Z^+$
coincides with the above number. \\

\vspace{0.5cm}
\underline{$T=\frac 12,\; S=0$ state (the $N$ resonance)}

\bear
\Gamma(N_{\at} \rightarrow N\pi)&=&
\frac{3 G_{\at}^2}{2\pi (M_N+M_{\at})^2 } |\vec{p}|^3 \frac{M_N}{M_{\at}}
\label{w1}
\cdot\frac{1}{20}\bigl(1-\frac{23}{2 \sqrt 5} c_{\at}\bigr)
=5\,MeV, \\
\Gamma(N_{\at} \rightarrow N\eta)&=&
\frac{3 G_{\at}^2}{2\pi (M_N+M_{\at})^2 } |\vec{p}|^3 \frac{M_N}{M_{\at}}
\cdot\frac{1}{20}\bigl(1-\frac{1}{2 \sqrt 5} c_{\at}\bigr)
=11\,MeV, \\
\Gamma(N_{\at} \rightarrow \Delta\pi)&=&
\frac{3 G_{\at}^2}{2\pi (M_\Delta+M_{\at})^2 } |\vec{p}|^3
\frac{M_\Delta}{M_{\at}} \cdot\frac{4}{5} c_{\at}^2
=5\,MeV,
\label{w3} \\
\Gamma(N_{\at}
\rightarrow \Lambda K)&=&
\frac{3 G_{\at}^2}{2\pi(M_\Lambda+M_{\at})^2 }|\vec{p}|^3
\frac{M_\Lambda}{M_{\at}} \cdot\frac{1}{20}\bigl(1+\frac{8}{ \sqrt 5}
c_{\at}\bigr)
=5\,MeV, \\
\Gamma(N_{\at} \rightarrow \Sigma K)&=& \frac{3
G_{\at}^2}{2\pi (M_\Sigma+M_{\at})^2}
|\vec{p}|^3 \frac{M_\Sigma}{M_{\at}}
\cdot\frac{1}{20}\bigl(1+\frac{3}{ \sqrt 5} c_{\at}\bigr)
=0.5\,MeV,
\label{w5}
\ear

These partial widths sum up into 27.5 MeV. However, the quantum numbers
of $N(1710)$ allow decays into, say, $N\pi\pi$ states which are not
fully accounted for above. Allowing a 50\% branching ratio for the
non-accounted decays, we estimate the full width as
$\Gamma_{tot}(N_{\at})\approx 27.5\,MeV\cdot 1.5 \approx 41\,MeV$,
where from the branching ratios can be deduced, see Table~3.

\vspace{.5cm}\hspace{1cm}
\underline{$T=1,\;  S=-1$ state (the $\Sigma$ resonance)}

\bear
\Gamma(\Sigma_{\at} \rightarrow N\bar K)&=&
\frac{3 G_{\at}^2}{2\pi (M_N+M_{\Sigma_{\at}})^2 }
|\vec{p}|^3 \frac{M_N}{M_{\Sigma_{\at}}}
\label{ws1}
\cdot\frac{1}{30}\bigl(1-\frac{9}{ \sqrt 5} c_{\at}\bigr)
=6\,MeV, \\
\Gamma(\Sigma_{\at} \rightarrow \Sigma\pi)&=&
\frac{3 G_{\at}^2}{2\pi (M_\Sigma+M_{\Sigma_{\at}})^2 }
|\vec{p}|^3 \frac{M_\Sigma}{M_{\Sigma_{\at}}}
\cdot\frac{1}{30}\bigl(1- \sqrt 5 c_{\at}\bigr)
=10\,MeV,
\\
\Gamma(\Sigma_{\at} \rightarrow \Sigma\eta)&=&
\frac{3 G_{\at}^2}{2\pi (M_\Sigma+M_{\Sigma_{\at}})^2 }
|\vec{p}|^3 \frac{M_\Sigma}{M_{\Sigma_{\at}}}
\cdot\frac{1}{20}\bigl(1+\frac{6}{ \sqrt 5} c_{\at}\bigr)
=9\,MeV,
\label{ws3} \\
\Gamma(\Sigma_{\at}
\rightarrow \Lambda \pi)&=&
\frac{3 G_{\at}^2}{2\pi(M_\Lambda+M_{\Sigma_{\at}})^2 }|\vec{p}|^3
\frac{M_\Lambda}{M_{\Sigma_{\at}}}
\cdot\frac{1}{20}\bigl(1-\frac{6}{ \sqrt 5}
c_{\at}\bigr)
=17\,MeV, \\
\Gamma(\Sigma_{\at}
\rightarrow \Xi K)&=&
\frac{3 G_{\at}^2}{2\pi(M_\Xi+M_{\Sigma_{\at}})^2 }|\vec{p}|^3
\frac{M_\Xi}{M_{\Sigma_{\at}}}
\cdot\frac{1}{30}\bigl(1+\frac{19}{ \sqrt 5}
c_{\at}\bigr)
=3 \,MeV, \\
\Gamma(\Sigma_{\at} \rightarrow \Sigma^* \pi)&=& \frac{3
G_{\at}^2}{2\pi (M_{\Sigma^*}+M_{\Sigma_{\at}})^2}
|\vec{p}|^3 \frac{M_{\Sigma^*}}{M_{\Sigma_{\at}}}
\cdot\frac{2}{15} c_{\at}^2
=2\,MeV.
\label{ws5}
\ear

These partial widths sum up to $47\,MeV$. Multiplying it by a factor of
1.5 as in the previous case we estimate the full width of the
$\Sigma_{\at}$ to be $\Gamma_{tot}(\Sigma_{\at})\approx 70\,MeV$. See
Table~4 for the calculated branching ratios and a comparison with the
data.

\vspace{.5cm}\hspace{1cm}
\underline{$T=3/2,\;  S=-2$ state (the exotic $\Xi_{3/2}$ baryon)}

\bear
\Gamma(\Xi_{3/2}\rightarrow \Sigma K) &=& 52\,MeV, \\
\Gamma(\Xi_{3/2}\rightarrow \Xi \pi) &=& 42\,MeV,\\
\Gamma_{tot}(\Xi_{3/2})&\approx& 140\,MeV.
\la{wx2}\ear

\vskip 0.33cm
\begin{tabular}{|c|c|c|}
\hline
\hline
$N_{\at}$ & prediction & data \\
\hline
$M_{\at}$\,, MeV & $1710$ {\rm(input)} & $1710$  \\
$\Gamma_{tot}(N_{\at})$\,, MeV & $\sim 40$ & 50 to 250\\
$Br(N\pi)$ &$\sim 0.13$ & 0.10 to 0.20\\
$Br(N\eta)$ &$\sim 0.28$ & $0.16\pm 0.10$ \\
$Br(\Delta \pi)$ P-wave &$\sim 0.13$ & --\\
$Br(\Lambda K)$ &$\sim 0.13$ & -- \\
$Br(\Sigma K)$ &$\sim 0.01$ & --\\
$\sqrt{Br(N\pi)Br(N\eta)}$ &$\sim 0.19$ &$0.30\pm 0.08$ \\
$\sqrt{Br(N\pi)Br(\Lambda K)}$ &$\sim 0.13$ &0.12 to 0.18 \\
$\sqrt{Br(N\pi)Br(\Delta \pi)}$ &$\sim 0.12$ &0.16 to 0.22 \\
\hline
\hline
\end{tabular} \vskip 0.5cm

Table 3. Predictions for decay modes of $N_{\at}$ identified with
$N\left(1710, \frac{1}{2}^+\right)$, confronted with the data from \cite{PDG}.
\vskip 0.33cm

Despite the smallness of the octet--anti-decuplet mixing represented by
the coefficient $c_{\at}$ (see \eq{mixnum}) it has a large impact on the
decay widths of the anti-decuplet because the decay channels
$8\rightarrow 8+8$ and $8\rightarrow 10+8$ are enhanced ``kinematically"
by large Clebsch--Gordan coefficients. For example, without taking into
account this mixing, the decay $N_{\at}\rightarrow \Delta \pi$ is forbidden,
however the small mixing probability, $c_{\at}^2 \sim 0.007$,
is amplified by a huge ``kinematical" factor $\sim 20$.

Finally, we mention that the $Z^+NK$ coupling correspondent to
\eqs{meson-soliton}{meson-soliton-rot}
can be written down in a relativistically-invariant
form as

\beq
L_{int}=ig_{KNZ}\left[(\bar p\gamma_5 Z^+)\bar K^0 + (\bar n \gamma_5
Z^+)K^- \right]\,.
\la{relinv}\eeq
Comparing its non-relativistic
limit with particular projections of
\eqs{meson-soliton}{meson-soliton-rot} we find

\beq
g_{KNZ}=\frac{3}{\surd{30}}\;\frac{2m_N}{m_N+m_Z}\;(G_0-G_1)\approx
4.1\,.  \la{gKNZ}\eeq

For a comparison, the ordinary $\Sigma^+NK$ vertex written in the form of
\eq{relinv} corresponds to $g_{KN\Sigma^+}\approx 5$.

\section{Identification of members of the anti-decuplet}

We see that the predicted branching ratios and total width of the $N(1710)$
are in a reasonable
agreement with the data, given the large errors and inconsistencies in the
data, see \cite{PDG}. Our numbers should be compared also with the
predictions for the $N(1710)$ decays following from the standard $SU(6)$
quark model, performed in ref. \cite{FC}. Assuming $N(1710)$ to be a member
of a normal octet the authors get, in particular, $\Gamma(N\eta) \approx
\Gamma(\Lambda K) \approx 0$, which seems to contradict the data
even though the errors are large. It should be mentioned, however, that
a recent analysis \cite{Batinic} suggests that in the $\sim 1700\,MeV$
region there might be actually two nucleon resonances: one coupled
stronger to pions and another to the $\eta$ meson.

We conclude that the $N(1710)$ nucleon resonance
is a good candidate for the $N_{\at}$ member of the anti-decuplet. Let us
stress that the octet--anti-decuplet mixing is important for the analysis.
It leads to a considerable reduction of the $N \pi$
branching ratio and of the total width;
simultaneously a non-zero $\Delta \pi$ branching ratio appears, in
accordance with the phenomenology of the $N(1710)$ decays.

There is a fair candidate for the $\Sigma_{\at}$ member of the anti-decuplet,
namely the $\Sigma(1880)$ from the Particle Data Group baryon listings
\cite{PDG}. The resonance has only a two-star status, and its properties are
not measured properly, including the mass ranging from $1826\pm 20$ to
$1985\pm 50$ MeV. Nevertheless, we compare our predictions for the
$\Sigma_{\at}$ with what is known about the resonance, see Table~4.

\vskip 0.33cm
\begin{tabular}{|c|c|c|}
\hline
\hline
$\Sigma_{\at}$ & prediction & data \\
\hline
$M_{\Sigma_{\at}}$\,, MeV
& $1890$ & $\approx 1880$  \\
$\Gamma_{tot}(\Sigma_{\at})$\,, MeV & $\sim 70$ &$80$ to 250\\
$Br(N\bar K)$ &$\sim 0.09$ & 0.06 to 0.3\\
$\sqrt{Br(N\bar K)Br(\Sigma\pi)}$ &$\sim 0.11$ &$\sim 0.3$ \\
$\sqrt{Br(N\bar K)Br(\Lambda\pi)}$ &$\sim 0.15$ &0.11 to 0.25\\
\hline
\hline
\end{tabular} \vskip 0.5cm

Table 4. Predictions for decay modes of $\Sigma_{\at}$, confronted with
the data from \cite{PDG}.
\vskip 0.33cm

What can be said is that the suggested identification does not contradict
the (poor) data on the $\Sigma(1880)$ resonance.

As to $\Xi_{3/2}$ which we predict at $2070\,MeV$, there are several
candidates for the non-exotic components of this quadruplet in the range
of masses between $1900$ and $2100\,MeV$, however even their quantum numbers
are not well established yet. In view of the estimate that our $\Xi_{3/2}$ is
wider than $140\,MeV$  it would be quite difficult to pinpoint
such a state, including its exotic components, $\Xi^{--}$ and $\Xi^+$.
Moreover, a presence of such a wide state would seriously influence the
determination of the parameters of other $\Xi$-type resonances, were they to
appear in this mass region. We sum up our predictions for the
anti-decuplet in Table~5.

\vskip 0.33cm
\begin{tabular}{|c|c|c|c|c|c|}
\hline
\hline
 &T&Y &Mass in MeV & Width in MeV & Possible candidate \\
\hline
$Z^+$&$0$ &$2$ & 1530 & 15 & ---\\
$N_{\at} $&$1/2$&$1$ & $1710\;{\rm(input)}$ & $\sim 40$ &
$N(1710)P_{11}$\\ $\Sigma_{\at}$&$1$ &$0$ & 1890 & $\sim 70$&
$\Sigma(1880)P_{11}$\\
$\Xi_{3/2}$ & $3/2$ & $-1$ & 2070 & $>140$ &
$\Xi(2030)?$ \\
\hline \hline
\end{tabular} \vskip 0.5cm

Table 5. Predictions for masses and total widths of the members of the
anti-decuplet and possible candidates for these states.
\vskip 0.33cm

It should be mentioned that the masses of the anti-decuplet have been
estimated in the Skyrme model with the results ranging from
$M_{Z^+}\approx 1500$~MeV \cite{Prasz2} to $\approx 1700$~MeV
\cite{Walliser}. Such an uncertainty arises in the Skyrme model
since one has to make a difficult choice between having the
nucleon mass correct and the $F_\pi$ constant wrong, or {\em vice
versa}. Predictions for the exotic $T=0, S=1$ $P_{01}$ state
in the bag model are grouped around 1750~MeV \cite{corden,bag}, that is
significantly higher than our estimate.

Our considerations have essentially been based on the identification of the
non-exotic member of the anti-decuplet with the rather well established
nucleon resonance, $N\left(1710,\frac{1}{2}^+\right)$. On general grounds,
we cannot exclude the possibility that the anti-decuplet as a whole
lies higher.  For example, the exotic $Z^+$ might, in principle, have a
mass of more than 1750 MeV (lower masses are probably excluded by the
old phase-shift analyses -- see ref. \cite{Dover}). There have been
claims in the past for observing such states (see ref.\cite{obzor} for
a review). In this case, however, there would be difficulties in
finding an appropriate candidate for the $N_{\at}$ member of the
anti-decuplet.  The only possibility suggested by the Particle Data
baryon listings is the $N\left(2100,\frac 12^+\right)$ resonance.
In this case the moment of inertia $I_2$ determining the shift of the
anti-decuplet center in respect to the decuplet center would be
very small (about $\sim 0.3$~fm).  Such a small moment of inertia can
hardly be obtained in any dynamical realization of the idea of baryons
as solitons, which seems to be so successful everywhere else.
Therefore, we believe that we present a good case for a relatively
light and narrow exotic baryon: it probably has not been observed in
the past just for these reasons.

\section{Conclusions}

The chiral soliton models of baryons, which correctly emphasize the
important role of the spontaneous chiral symmetry breaking in the dynamics of
strong interactions, are extremely successful in explaining relations between
octet and decuplet baryons since in these models they all appear as various
rotational excitations of the same object.

The two lowest rotational states
of chiral solitons are exactly the octet with spin $\frac{1}{2}$ and the
decuplet with spin $\frac{3}{2}$, and it is natural to ask oneself what
is the next rotational state. The answer is
\cite{DP,Chemtob,Prasz2,Walliser}: it is
the anti-decuplet with spin $\frac{1}{2}$, and most of its properties
can be predicted from symmetry considerations only, without entering into
dynamics which is model-dependent. The only unknown parameter (a specific
$SU(3)$ moment of inertia) can be fixed by identifying the nucleon-like
member of the anti-decuplet with the observed
$N\left(1710,\frac{1}{2}^+\right)$
resonance. Its decay modes, as well as masses and decay modes of the
other members of the anti-decuplet can be then fixed unambigiously.
The calculated decay modes of the $N(1710)$ are found to be in a reasonable
agreement with the existing data though the data are not good enough
to make a decisive conclusion.
At least it seems that the standard non-relativistic $SU(6)$
description of this state as a member of an octet, is in trouble with the
data: the anti-decuplet idea fits better.

In a sense, history repeats itself: there are candidates for all members
of the anti-decuplet, except for its vertex -- like in the early 60's when
all members of the now venerable decuplet were known except the
$\Omega^-$ hyperon. In our case it is the exotic $Z^+$ baryon, which
decays into $K^+\,n$ and $K^0\,p$.  Claims for observing such states
have been made in the past (see ref.\cite{obzor} for a review) but they
are all substantially higher than our prediction $m_Z\approx 1530\,MeV$,
with the width lower than $\Gamma_Z\approx 15\,MeV$.

The most direct way to detect the exotic $Z^+$ resonance would be in the
$K^0\,p$ or $K^+\,n$ scattering. Unfortunately, the mass range in question
is too low for kaon beams and probably too high for the $\phi$
factory of kaons.

Another possibility to reveal the exotic $Z^+$ is in the
collisions of non-strange particles. In comparison with direct
production in $KN$ collision, such reactions are more complicated as they
involve many particles in the final states of which
some are neutral and some are charged, therefore a combined detector is
needed. As to the missing-mass-type experiments they seem to be vulnerable
because of severe background conditions and the narrowness of the
$Z^+$.  Let us list several possibilities of the $Z^+$ production in
reactions with non-strange particles.

\begin{itemize}
\item
\underline{Nucleon--nucleon collisions}\\
$pn \to \Lambda Z^+ \to \Lambda K^+ n \; \mbox{or}\; \Lambda K^0 p$,
$\;p_{lab} > 2.60 \;GeV/c$ \\
$pp \to \Sigma^+ Z^+ \to \Sigma^+ K^+ n \; \mbox{or}\; \Sigma^+ K^0 p$,
$\;p_{lab}>2.8\; GeV/c$
\item
\underline{Photon-nucleon collisions}\\
$\gamma p \to \bar K^0 Z^+ \to  \bar K^0 K^+ n
\; \mbox{or}\; \bar K^0 K^0 p$, $\;p_{lab}>1.7 \;GeV/c$ \\
$\gamma n \to K^-Z^+ \to K^- K^+ n
\; \mbox{or}\; K^-K^0 p$, $\;p_{lab}>1.7 \;GeV/c$
\item
\underline{Pion-nucleon collisions}\\
$\pi^- p \to K^-Z^+ \to K^- K^+ n
\; \mbox{or}\; K^-K^0 p$ $\;p_{lab}>1.7 \;GeV/c$ \\
$\pi^+ n \to \bar K^0Z^+\to \bar K^0K^+n
\; \mbox{or}\; \bar K^0K^0p$, $\;p_{lab}>1.7 \;GeV/c$.
\end{itemize}

One of the most promising ways to check the existence of the
$Z^+$ baryon would be in the photon collisions with energies
$>2\,GeV$, since the photon already carries a portion of strange quarks
\footnote{One of us (D.D.) is grateful to E.Paul for a conversation on
this point.}.  Another possibility is to analyze the LASS data
from the 11 GeV $K^+ p$ collisions \footnote{We thank J.~Bjorken 
for this suggestion.}.
In any case, a search for a light and narrow exotic $Z^+$ baryon
seems to be a challenging task.

\section{Acknowledgments}
This work has been supported in part by the INTAS grants 93-0283 EXT
and 93-1630-EXT and by a joint grant
of the Deutsche Forschungsgemeinschaft and the Russian Foundation for
Basic Research. M.P. is a
A.v.Humboldt Fellow, and acknowledges warm hospitality by Prof. Klaus
Goeke at the Bochum University.

We thank P.Pobylitsa for a cooperation. M.P. is grateful
to M. Praszalowicz and H.~Walliser for useful discussions.

\setcounter{section}{0}

\appendix
\section{A}
The rotational wave functions of baryons are eigenfunctions of the collective
hamiltonian

\[
H=H^{rot}+\Delta H_m,
\]
where the $SU(3)$-symmetric $H^{rot}$ is given by \eq{Ham1} and
the $SU(3)$-breaking part, $\Delta H_m$, is given by \eq{Ham2}.
The eigenfunctions of the unperturbed hamiltonian $H^{rot}$ are
proportional to the Wigner finite-rotation matrices \cite{DS}:

 \beq
 |B\rangle=|B, r\rangle=\sqrt{\dim \;r} (-1)^{J_3-1/2} D^{(\bar
r)}_{Y,T,T_3;1,J,-J_3},
\label{wave-functions}
\eeq
 where $r$ is an irreducible representation of the
 $SU(3)$ group, $r=8,10,\at$, etc., $B$ denotes a set of quantum
 numbers: $Y,T,T_3$ (hypercharge, isospin and its projection) and $J,
 J_3$ (spin and its projection).

 In the linear approximation in $m_s$ rotational wave functions are
 superpositions of different representations:

 \beq
 |B\rangle=|B, r \rangle+\sum_{r'\neq r }
 \frac{\langle B,r'|\Delta H_m|B,r\rangle}{E_B^{r(0)}-E_B^{r'(0)}}|B,r'\rangle.
 \label{perth}
 \eeq
 Here the unperturbed energies $E_B^{r(0)}$ are given by \eq{rotenerg}.
Neglecting
admixtures of the 27- and 35-plets and using the general \eq{perth} we
obtain the wave functions (\ref{first}-\ref{last}).

\appendix
\section{B}

In this Appendix we derive general formulae for the decay rate of a
baryon $B_1$
with flavour quantum numbers $(YTT_3)=(Y_1 T_1 t_1)$ and spin
$(JJ_3)=(J_1j_1)$, into a baryon $B_2$
with quantum numbers $(Y_2 T_2 t_2)$ and
$(J_2j_2)$, plus an octet pseudoscalar meson with $(YTT_3)=(Y_m T_m t_m)$.
In order to obtain the amplitude one has to sandwich the meson-soliton
coupling
\beq
-i \frac{3 G_0}{2 M_B}\cdot
\frac 12 \Tr(R^\dagger\lambda^m R\lambda^i)\cdot p_i,
\label{meson-soliton-app}
 \eeq
where $\lambda^m$ is the Gell-Mann matrix of the
correspondent meson, $R$ is the matrix describing the orientation of the
soliton, and $p_i$ is the 3-momentum of the meson, between rotational
wave functions describing the baryons $B_2$ and $B_1$. To incorporate a
general situation we assume that their rotational wave
functions are mixtures of certain $SU(3)$ muiltiplets $r$ and $q$, so
that one can write the wave functions as linear
combinations of the Wigner $D$-functions:

\beq
\Psi_{B_i}(R)= (-1)^{j_i-1/2} \bigl\{
\sqrt{\dim \;r_i} D^{(\bar
r_i)}_{Y_i,T_i,t_i;1,J_i,-j_i}  +
A_i
\sqrt{\dim \;q_i} D^{(\bar
q_i)}_{Y_i,T_i,t_i;1,J_i,-j_i} \bigr\}.
\eeq
We assume the admixtures $A_{1,2}$ to be small ($\sim m_s$), and
neglect systematically the $A_i^2$ terms.

Using the fact that

\beq
\frac 12 \Tr(R^\dagger\lambda^m R\lambda^i) = D^{(8)}_{m,i}(R),
\eeq
and the general formula,

\[ \int dR D^{r*}_{\nu\nu^\prime}(R)D^{r_1}_{\nu_1\nu^\prime_1}(R)
D^{r_2}_{\nu_2\nu_2^\prime}(R) \]
\beq =\frac{1}{dim(r)}\sum_{\mu^\prime}\delta_{rr^\prime}
\left( \begin{array}{ccc}
r_1 & r_2 & r^\prime \\ \nu_1 & \nu_2 &  \nu
\end{array} \right)
\left( \begin{array}{ccc}
r_1 & r_2 & r^\prime \\ \nu_1^\prime &\nu_2^\prime & \nu^\prime,
\end{array} \right),
\eeq
where the sum goes over all occurrences of the representation $r$ in the
product of representations $r_1$ and $r_2$, one gets for the decay
amplitude $B_1\to B_2 +M$ the following expression in terms of the $SU(3)$
Clebsch--Gordan coefficients :

\bear
\nonumber
&&{\cal M}(B_1\to B_2 M) = \frac{3 g}{M_2+M_1}p^i
\\
\nonumber
&&\Biggl\{
\sqrt{\frac{r_2}{ r_1}} \sum_{r_1'}
\left(
\begin{array}{ccc}
r_2 & 8 & r_1'\\
Y_2 T_2 t_2 &  Y_mT_mt_m & Y_1 T_1 t_1
\end{array}\right)
\left(
\begin{array}{ccc}
r_2 & 8 & r_1'\\
1 J_2 j_2 &  01i & 1 J_1 j_1
\end{array}\right) +\\
\nonumber
&& A_1
\sqrt{\frac{r_2}{ q_1}} \sum_{q_1'}
\left(
\begin{array}{ccc}
r_2 & 8 & q_1'\\
Y_2 T_2 t_2 &  Y_mT_mt_m & Y_1 T_1 t_1
\end{array}\right)
\left(
\begin{array}{ccc}
r_2 & 8 & q_1'\\
1 J_2 j_2 &  01i & 1 J_1 j_1
\end{array}\right)+\\
\nonumber
&& A_2
\sqrt{\frac{q_2}{ r_1}} \sum_{r_1'}
\left(
\begin{array}{ccc}
q_2 & 8 & r_1'\\
Y_2 T_2 t_2 &  Y_mT_mt_m & Y_1 T_1 t_1
\end{array}\right)
\left(
\begin{array}{ccc}
q_2 & 8 & r_1'\\
1 J_2 j_2 &  01i & 1 J_1 j_1
\end{array}\right)
\Biggr\}.
\ear
Before we square this amplitude let us factorize out the dependence
on the $SU(2)$ quantum numbers (referring both to spin and isospin)  using
the relation between the $SU(3)$ and the $SU(2)$ Clebsch--Gordan coefficients
\cite{DS} (the proportionality coefficient is called the isoscalar factor):

\beq
\left(
\begin{array}{ccc}
r_1 & r_2 & r_3\\
Y_1 T_1 t_1 &  Y_2T_2t_2 & Y_3 T_3 t_3
\end{array}\right)
= C^{T_3 t_3}_{T_1 t_1; T_2 t_2}
\left(
\begin{array}{cc}
r_1 & r_2 \\
Y_1 T_1 & Y_2 T_2
\end{array}\right|\left.\begin{array}{c}
		  r_3\\Y_3T_3
		  \end{array}\right).
\eeq
Making use of the above relation one gets for the amplitude squared:

\bear
\nonumber
&&|{\cal M}|^2 = \frac{9 G_0^2}{(M_2+M_1)^2}p^i p^j
|C^{T_1 t_1}_{T_2,t_2;T_m ,t_m}|^2
C^{J_1 j_1}_{J_2,j_2;1i }
C^{J_1 j_1}_{J_2,j_2;1j }
\\
&& \times\Biggl\{
\nonumber
\frac{r_2}{r_1}\Biggl|
\sum_{r_1'}
\left(
\begin{array}{cc}
r_2 & 8 \\
Y_2 T_2 & Y_m T_m
\end{array}\right|\left.\begin{array}{c}
		  r_1'\\Y_1 T_1
		  \end{array}\right)
\left(
\begin{array}{cc}
r_2 & 8 \\
1 J_2 & 0 1
\end{array}\right|\left.\begin{array}{c}
		  r_1'\\1 J_1
		  \end{array}\right)
\Biggr|^2 +\\
&&
\nonumber
2 A_1 \frac{r_2}{\sqrt{q_1 r_1}}
\sum_{r_1'}
\left(
\begin{array}{cc}
r_2 & 8 \\
Y_2 T_2 & Y_m T_m
\end{array}\right|\left.\begin{array}{c}
		  r_1'\\Y_1 T_1
		  \end{array}\right)
\left(
\begin{array}{cc}
r_2 & 8 \\
1 J_2 & 0 1
\end{array}\right|\left.\begin{array}{c}
		  r_1'\\1 J_1
		  \end{array}\right)\\
 &&\times\sum_{q_1'}
\label{msq}
\left(
\begin{array}{cc}
r_2 & 8 \\
Y_2 T_2 & Y_m T_m
\end{array}\right|\left.\begin{array}{c}
		  q_1'\\Y_1 T_1
		  \end{array}\right)
\left(
\begin{array}{cc}
r_2 & 8 \\
1 J_2 & 0 1
\end{array}\right|\left.\begin{array}{c}
		  q_1'\\1 J_1
		  \end{array}\right)
 +\\
&&
\nonumber
2 A_2 \frac{\sqrt{r_2 q_2}}{ r_1}
\sum_{r_1'}
\left(
\begin{array}{cc}
r_2 & 8 \\
Y_2 T_2 & Y_m T_m
\end{array}\right|\left.\begin{array}{c}
		  r_1'\\Y_1 T_1
		  \end{array}\right)
\left(
\begin{array}{cc}
r_2 & 8 \\
1 J_2 & 0 1
\end{array}\right|\left.\begin{array}{c}
		  r_1'\\1 J_1
		  \end{array}\right)\\
&&\times\sum_{r_1'}
\nonumber
\left(
\begin{array}{cc}
q_2 & 8 \\
Y_2 T_2 & Y_m T_m
\end{array}\right|\left.\begin{array}{c}
		  r_1'\\Y_1 T_1
		  \end{array}\right)
\left(
\begin{array}{cc}
q_2 & 8 \\
1 J_2 & 0 1
\end{array}\right|\left.\begin{array}{c}
		  q_1'\\1 J_1
		  \end{array}\right)
\Biggr\}.
\ear
To get the decay width one needs to average the amplitude
squared $|{\cal M}|^2$ over the initial and to sum over the final spin
and isospin states:

\beq
\overline{{\cal M}^2} = \frac{1}{(2 T_1+1)(2 J_1+1)}
\sum_{t_2 t_m j_2} \sum_{t_1 j_1}  |{\cal M}|^2.
\eeq
In eq.~(\ref{msq}) the dependence on spin and isospin projections is
factored out, hence one can perform the summation over final and initial
spin and isospin states with the help of the orthogonality relations for the
$SU(2)$ Clebsh--Gordan coefficients:

\bear
&&\sum_{j_2j_1 }
C^{J_1 j_1}_{J_2,j_2;1i }
C^{J_1 j_1}_{J_2,j_2;1j } =\frac{2 J_1+1}{3} \delta_{ij}, \\
&&\sum_{t_2t_1t_m }
|C^{T_1 t_1}_{T_2,t_2;T_m ,t_m}|^2 =2T_1+1.
\ear
Multiplying $\overline{{\cal M}^2}$ by the phase volume we
get the final result for the decay rate of $B_1 \to B_2 +M$
in terms of the $SU(3)$ isoscalar factors:

\bear
\nonumber
&&\Gamma(B_1\to B_2+M) = \frac{3 G_0^2}{2\pi (M_2+M_1)^2}|p|^3
\frac{M_2}{M_1}
\\ &&\times \Biggl\{ \nonumber \frac{r_2}{r_1}\Biggl|
\sum_{r_1'}
\left(
\begin{array}{cc}
r_2 & 8 \\
Y_2 T_2 & Y_m T_m
\end{array}\right|\left.\begin{array}{c}
		  r_1'\\Y_1 T_1
		  \end{array}\right)
\left(
\begin{array}{cc}
r_2 & 8 \\
1 J_2 & 0 1
\end{array}\right|\left.\begin{array}{c}
		  r_1'\\1 J_1
		  \end{array}\right)
\Biggr|^2 +\\
&&
\nonumber
2 A_1 \frac{r_2}{\sqrt{q_1 r_1}}
\sum_{r_1'}
\left(
\begin{array}{cc}
r_2 & 8 \\
Y_2 T_2 & Y_m T_m
\end{array}\right|\left.\begin{array}{c}
		  r_1'\\Y_1 T_1
		  \end{array}\right)
\left(
\begin{array}{cc}
r_2 & 8 \\
1 J_2 & 0 1
\end{array}\right|\left.\begin{array}{c}
		  r_1'\\1 J_1
		  \end{array}\right)\\
 &&\times\sum_{q_1'}
\label{rate}
\left(
\begin{array}{cc}
r_2 & 8 \\
Y_2 T_2 & Y_m T_m
\end{array}\right|\left.\begin{array}{c}
		  q_1'\\Y_1 T_1
		  \end{array}\right)
\left(
\begin{array}{cc}
r_2 & 8 \\
1 J_2 & 0 1
\end{array}\right|\left.\begin{array}{c}
		  q_1'\\1 J_1
		  \end{array}\right)
 +\\
&&
\nonumber
2 A_2 \frac{\sqrt{r_2 q_2}}{ r_1}
\sum_{r_1'}
\left(
\begin{array}{cc}
r_2 & 8 \\
Y_2 T_2 & Y_m T_m
\end{array}\right|\left.\begin{array}{c}
		  r_1'\\Y_1 T_1
		  \end{array}\right)
\left(
\begin{array}{cc}
r_2 & 8 \\
1 J_2 & 0 1
\end{array}\right|\left.\begin{array}{c}
		  r_1'\\1 J_1
		  \end{array}\right)\\
&&\times\sum_{r_1'}
\nonumber
\left(
\begin{array}{cc}
q_2 & 8 \\
Y_2 T_2 & Y_m T_m
\end{array}\right|\left.\begin{array}{c}
		  r_1'\\Y_1 T_1
		  \end{array}\right)
\left(
\begin{array}{cc}
q_2 & 8 \\
1 J_2 & 0 1
\end{array}\right|\left.\begin{array}{c}
		  q_1'\\1 J_1
		  \end{array}\right)
\Biggr\}.
\ear
These $SU(3)$ isoscalar factors can be found in ref. \cite{DS}.

Similar derivation can be performed for the next-to-leading pseudoscalar
couplinigs eqs.~(\ref{meson-soliton-rot}).
The results are summarized in Table~2.

\newpage
\newpage
\setlength{\unitlength}{0.00073300in}%
\begingroup\makeatletter\ifx\SetFigFont\undefined
\def\x#1#2#3#4#5#6#7\relax{\def\x{#1#2#3#4#5#6}}%
\expandafter\x\fmtname xxxxxx\relax \def\y{splain}%
\ifx\x\y   
\gdef\SetFigFont#1#2#3{%
  \ifnum #1<17\tiny\else \ifnum #1<20\small\else
  \ifnum #1<24\normalsize\else \ifnum #1<29\large\else
  \ifnum #1<34\Large\else \ifnum #1<41\LARGE\else
     \huge\fi\fi\fi\fi\fi\fi
  \csname #3\endcsname}%
\else
\gdef\SetFigFont#1#2#3{\begingroup
  \count@#1\relax \ifnum 25<\count@\count@25\fi
  \def\x{\endgroup\@setsize\SetFigFont{#2pt}}%
  \expandafter\x
    \csname \romannumeral\the\count@ pt\expandafter\endcsname
    \csname @\romannumeral\the\count@ pt\endcsname
  \csname #3\endcsname}%
\fi
\fi\endgroup

\begin{figure}[h]
\begin{picture}(10224,8775)( 8000,-8173)
\makebox[8.520in]{\rule{0in}{-7.020in}
\thicklines
\put(7201,-5761){\circle*{150}}
\put(4801,-5761){\circle*{150}}
\put(2401,-5761){\circle*{150}}
\put(9601,-5761){\circle*{150}}
\put(8401,-3961){\circle*{150}}
\put(6001,-3961){\circle*{150}}
\put(3601,-3961){\circle*{150}}
\put(7201,-2161){\circle*{150}}
\put(4801,-2161){\circle*{150}}
\put(6001,-361){\circle*{150}}
\put(6001,-361){\line( 2,-3){3600}}
\put(4801,-2161){\line( 1,0){2400}}
\put(6001,-3961){\line( 2,3){1200}}
\put(6001,-3961){\line( -2,3){1200}}
\put(6001,-3961){\line( 1,0){2400}}
\put(6001,-3961){\line( -1,0){2400}}
\put(6001,-3961){\line( 2,-3){1200}}
\put(6001,-3961){\line( -2,-3){1200}}
\put(8401,-3961){\line( -2,-3){1200}}
\put(3601,-3961){\line( 2,-3){1200}}
\put(2401,-5761){\line( 1, 0){7200}}
\put(2401,-5761){\line( 2, 3){3600}}
\put(6451,-286){\makebox(0,0)[lb]{\smash{\SetFigFont{14}{24.0}{rm}
$Z^+ (1530)$}}}
\put(4000,-286){\makebox(0,0)[lb]{\smash{\SetFigFont{14}{24.0}{rm}
$
\begin{array}{c}
nK^+ \;\; \mbox{or} \;\;pK^0 \\
uudd\bar s
\end{array}
$
}}}
\put(7426,-2011){\makebox(0,0)[lb]{\smash{\SetFigFont{14}{24.0}{rm}
$N(1710)$
}}}
\put(8701,-3561){\makebox(0,0)[lb]{\smash{\SetFigFont{14}{24.0}{rm}
$\Sigma(1890)$
}}}
\put(1401,-6361){\makebox(0,0)[lb]{\smash{\SetFigFont{14}{24.0}{rm}
$
\begin{array}{c}
\Xi^- \pi^-  \;\; \mbox{or}\;\; \Sigma^- K^-
\\
dd s s \bar u
\end{array}$ }}}
\put(9151,-6361){\makebox(0,0)[lb]{\smash{\SetFigFont{14}{24.0}{rm}
$
\begin{array}{c}
\Xi^0 \pi^+  \;\; \mbox{or}\;\; \Sigma^+ \bar K^0
\\
uu s s \bar d
\end{array}$
}}}
\put(9551,-5361){\makebox(0,0)[lb]{\smash{\SetFigFont{14}{24.0}{rm}
$\Xi_{3/2}(2070)
$
}}}}
\end{picture}
\caption{ The suggested anti-decuplet of baryons. The corners of this
$(T_3, Y)$ diagram are exotic. We show their quark content together
with their (octet baryon+octet meson) content, as well as the predicted
masses.}
 \end{figure}
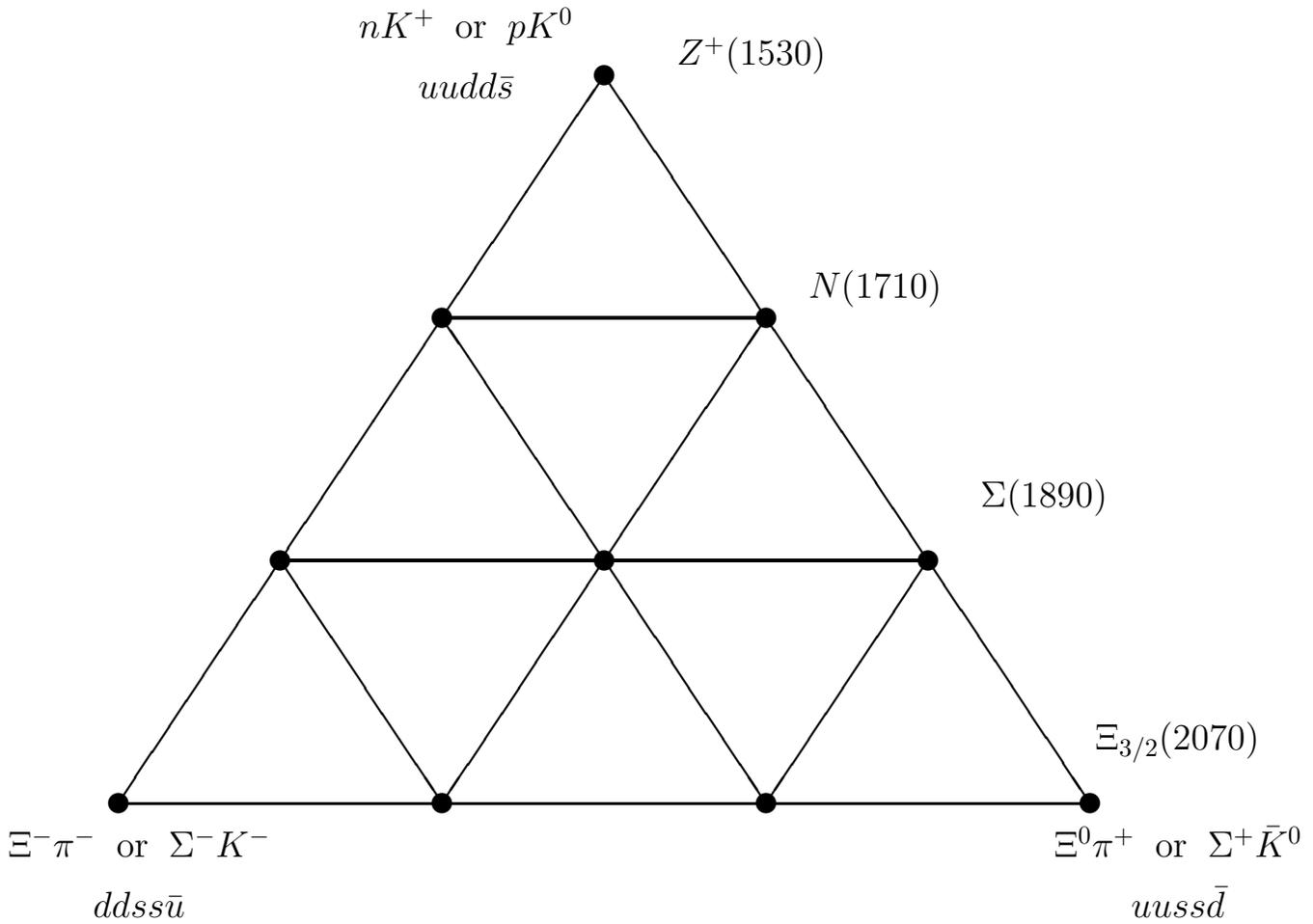
\end{document}